%% file: main.tex
\documentclass{article}
\usepackage{logic-article,amsmath,graphicx,pdfsync,tipa}
\usepackage{amssymb,amscd}
\usepackage[usenames,dvipsnames,svgnames,table]{xcolor}

\newcommand{\blank}{\_}
\newcommand{\lgderiv}[1]{g_{#1\blank}}
\newcommand{\lderiv}[1]{f_{#1\blank}}
\newcommand{\rderiv}[1]{f_{\blank#1}}
\newcommand{\mderiv}[2]{f_{#1\blank#2}}
\newcommand{\eqdef}{\stackrel {\mathrm{def}} =}

\newcommand{\markalpha}[1]{\textcolor {red}{#1}}

\newcommand{\leftvar}{\mbox{\rm \small left}}
\newcommand{\middlevar}{\mbox{\rm \small middle}}
\newcommand{\rightvar}{\mbox{\rm \small right}}
\newcommand{\leftblock}[1]{\overset{#1}\leftvar}
\newcommand{\middleblock}[1]{\overset{#1}\middlevar}
\newcommand{\rightblock}[1]{\overset{#1}\rightvar}

\newcounter{ourexamplecounter}
\newenvironment{ourexample}{
\medskip

\refstepcounter{ourexamplecounter}
\smallskip\noindent{\textbf{{Example \arabic{ourexamplecounter}. }}}}{\hfill $\Box$
\medskip 
}

\begin{document}
	\title{Transducers with origin information}
	\author{Miko{\l}aj Boja\'nczyk}
	\maketitle
	
	\begin{abstract}
		Call a string-to-string transducer regular if it can be realised by one of the following equivalent models:  mso transductions, two-way deterministic automata with output, and  streaming transducers with registers.
	This paper   proposes to treat origin information as part of the semantics of a regular string-to-string transducer. With such semantics, the model admits a machine-independent characterisation, Angluin-style learning in polynomial time, as well as effective characterisations of natural subclasses such as one-way transducers or first-order definable transducers.
		
	\end{abstract}

\input{intro}
\input{models}

\input{characterisation}

\input{learning}
\input{subclasses}
\input{fo}

\input{conclusion}

\bibliographystyle{alpha}
\bibliography{bib.bib}

\end{document}

%% file: intro.tex
This paper is about string-to-string transducers which, in one of several equivalent definitions,  can be described by deterministic two-way automata with output~\cite{DBLP:journals/jcss/AhoU70}. As shown in~\cite{DBLP:journals/tocl/EngelfrietH01}, this model is equivalent to  mso definable string transducers. Another equivalent model, used in~\cite{DBLP:conf/fsttcs/AlurC10}, is a  deterministic one-way automaton with registers that store parts of the output\footnote{Registers are similar to attributes in attribute grammars. The equivalence of  mso definable transductions with a form of attribute grammars, in the tree-to-tree case, was shown in~\cite{DBLP:journals/jcss/BloemE00}. In the special case of string-to-string transductions, the attribute grammars from~\cite{DBLP:journals/jcss/BloemE00} correspond to left-to-right deterministic automata with registers and regular lookahead.}.
Examples of such transducers include: duplication $w \mapsto ww$;  reversing $w \mapsto w^R$; a function $w \mapsto ww^R$ which maps an input to a palindrome whose first half is $w$; and a function which duplicates inputs of even length and reverses inputs of odd length.
As witnessed by the multiple equivalent definitions, this class of string-to-string transducers is robust, and therefore, following~\cite{DBLP:conf/fsttcs/AlurC10}, we call it  the class  of \emph{regular (string-to-string) transducers}. Regular transducers  have good closure properties. For instance, if $f$ and $g$ are regular, then  the composition $w \mapsto f(g(w))$ is also regular, which is straightforward if the mso definition is used, but nontrivial if the two-way automata  definition is used~\cite{DBLP:conf/icalp/ChytilJ77}. Also the   concatenation $w \mapsto f(w) \cdot g(w)$ is regular, which is aparent in any of the three definitions. Equivalence is of regular transducers is decidable, as was shown in~\cite{DBLP:journals/siamcomp/Gurari82} using the two-way automata definition.

\paragraph*{Origins.} The motivation of this paper is the simple observation that the models discussed above, namely determinsitic two-way automata with output, mso definable string transducers, and automata with registers, provide more than just a function from strings to strings. In each case, one can also reconstruct  \emph{origin information}, which  says how  positions of the output string originate from positions in the input string. How do we reconstruct the origin of a position $x$ in an output string? In the case of a deterministic two-way automaton, this is the position of the head when $x$ was output. In the case of an mso definable transducer, this is the position in which $x$ is interpreted. In the case of an automaton with registers, this is the position in the input when the letter $x$ was first loaded into a register.  In other colours, for a  transducer we can consider  two semantics: the \emph{standard semantics}, where the output is a string,  and the \emph{origin semantics}, where the output is a string with origin information. The second semantics is finer in the sense that  some transducers might be equivalent under the standard semantics, but not under the origin semantics.

Tracking origin information for transducers  has been studied before, for instance in the programming language community. Various tools for programming languages, such as evaluators, type checkers or translators, can be seen as tree-to-tree transducers, which transform syntax trees. The paper~\cite{DBLP:journals/jsc/DeursenKT93} provides a precise definition of origin information, and shows how it can be used visualize program execution, construct debuggers, and provide positional information in error reports. This idea is further developped in~\cite{DKT96.origins}, which in particular shows some results about  origin information for macro tree transducers, a powerful model that generalises the transducers considered in this paper.
Origin information has also been used as a technical tool in the study of tree-to-tree transducers. Examples include~\cite{DBLP:journals/siamcomp/EngelfrietM03}, where origin information is used to characterise those macro tree transducers which are mso definable, and \cite{DBLP:conf/pods/LemayMN10}, where origin information is used to get a Myhill-Nerode characterisation of deterministic top-down tree transducers. 

\paragraph*{Origin semantics.} 
To illustrate the difference between the two semantics (standard and origin) of a string-to-string transducer, consider a transducer which is the identity on the string $ab$, and which maps  other strings to the empty string. If we care about origins, then this description is incomplete. It could be that the first letter of the ouptut originates from the first letter of the input, and the second letter of the output corresponds to the second letter of the input (as realised by a two-way automaton, which first does a pass to determine if the input is $ab$, and then does a second pass which copies the input). It could also be that the whole output originates from the first letter in the input (as realised by a two-way automaton, which first does a pass to determine if the input is $ab$, and then returns to the first position where it outputs $ab$). Altogether, the same standard semantics can be described by four different origin semantics shown below.
\begin{center}
	\includegraphics[scale=0.3]{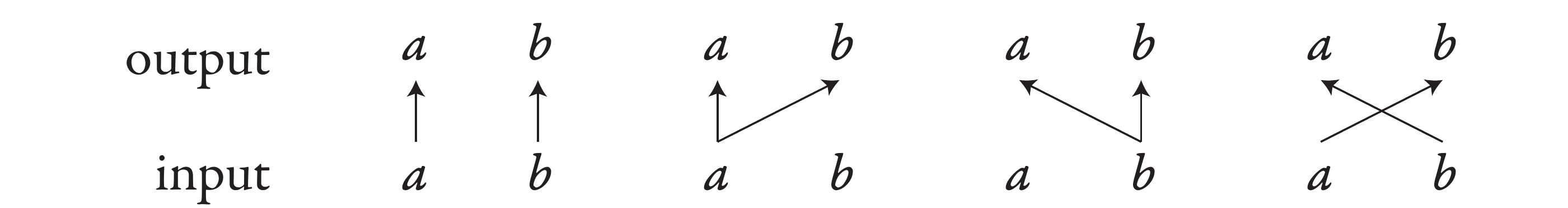}
\end{center}

Another example is the identity function on strings over a one letter alphabet $\set{a}$, which can be realised infinitely many different ways once origins are taken into account. One example is an automaton that outputs $a^n$  in input positions divisible by $n$, and then outputs the remainder under division by $n$ in the last input position.

This paper is a study of  the more refined semantics.  Almost any ``natural'' construction for transducers will respect origin information.  For instance, the translation from~\cite{DBLP:journals/tocl/EngelfrietH01} which converts an mso interpretation into a deterministic two-way automaton remains correct when the origin information is taken into account. The same holds for the other translations between the three models. In other colours, one can also talk about \emph{regular string-to-string transducers with origin information}. Various closure properties, such as composition and concatenation, are  retained when origins are taken into accout. Some results become easier to prove, e.g.~equivalence of string-to-string transductions.

\paragraph*{A machine independent characterisation.} The main contribution of this paper is a machine independent characterisation of regular transductions with origin information, which is given in Theorem~\ref{thm:myhill}. The characterisation is  similar to the Myhill-Nerode theorem, which says that a language $L$ is regular if and only if it has finitely many left derivatives of the form
\begin{align*}
	 w^{-1}L   \eqdef \set{v : wv \in L}.
\end{align*}
Furthermore, from the Myhill-Nerode theorem for regular languages  one obtains a canonical device, which is the minimal  deterministic automaton (a minimal right-to-left automaton is obtained if right derivatives are used, and a minimal monoid is obtained if two-sided derivatives are used).  The situation is similar for transducers with origin information. 
We define a notion of left and right derivatives for a transducer with origin information, and show that a transducer is regular if and only if it has finitely many left and right derivatives (finitely many left derivatives is not enough, same for right derivatives). The proof of the theorem yields a canonical device, which  is obtained from the function itself and not its representation as a two-way automaton, mso transduction, or machine with registers.  One use for the canonical device is testing equivalence: two devices are equivalent if and only if they yield the same canonical machine.

Another use of the canonical device is that it is  easy to see when the underlying function actually belongs to a restricted class, e.g.~if it can be defined by a deterministic  one-way automaton with output (see Theorem~\ref{thm:deterministic-without-lookahead}), or by functional nondeterministic one-way automaton with output (see Theorem~\ref{thm:order-preserving-characterisation}). A more advanced application is given in Theorem~\ref{thm:first-order transducer}, which characterises the firsto-order fragment of mso definable transducers with origin information.

\paragraph*{Learning.} One of the advantages of origin information is that it allows transducers to be learned, using an Angluin style algorithm. We show that  a transducer with origin information can be learned with a number of queries that is polynomial in the size of the canonical device. The  queries are of two types: the learner can ask for the value of the transducer on some input string; or the learner can propose a transducer with origin information, and in case this is not the correct one, then get a counterexample string where the proposed transducer produces a wrong output.

In the algorithm, the learner uses the origin information. However, it seems that the learner's advantage from the  origin information does not come at any signficant cost to the teacher.  Suppose that we want  to learn a transducer inside a text editor, e.g.~the user wants to teach the text editor that she is thinking of the  transducer which replaces every $=$ by $:=$.  If a user is trying to show an example of this transducer on some input, then she will probably place the cursor on occurrences of $=$ in the input, delete them, and retype $:=$, thus giving origin information to the algorithm. A user who  backspaces the whole input and retypes a new version will possibly be thinking of some different transformation. It would be wasteful to throw away this additional information supplied by the user.

\paragraph*{Thank you.} I would like to thank Sebastian Maneth for his valuable feedback; Anca Muscholl, Szymon Toru\'nczyk and Igor Walukiewicz for discussions about the model; and Rajeev Alur for asking the question about a machine-independent characterisation of transducers.


%% file: models.tex
	\section{Regular string to string transducers}
	
	A string-to-string transducer is any function from strings over some fixed input alphabet to strings over some fixed output alphabet. A string-to-string transducer with origin information is defined the same way, but for every input string $w$ it provides not only an output string $f(w)$, but also origin  information, which is a function from positions in $f(w)$ to positions in $w$.
	In this section we recall the definition of regular string-to-string transducers, and define the notion of regular string-to-string transducers with origin information. 
	Since we only consider string-to-string transducers in this paper, we simply say transducer from now on.
	
\paragraph*{Streaming transducer.}	
		An \emph{streaming transducer} is defined as follows. It has finite \emph{input} and  \emph{output alphabets}. There is a finite set of \emph{control states} with a distinguished \emph{initial state}, and  a finite set  of \emph{registers}, with a distinguished \emph{output register}. The transition function inputs a control state and an input letter, and outputs a new control state and  a \emph{register update}, which is  a sequence of register operations of two possible types:
		\begin{itemize}
			\item replace the contents of register $r$ with $rs$, and erase  register  $s$;
			\item replace the contents of register $r$ with output letter  $b$. 
		\end{itemize}
  Finally, there is an \emph{end of input function}, which maps each state to a sequence of register operations of the first type\footnote{The end of input function is prohibited to  produce new output letters so that  the origin information can be assigned. Alternatively, one could assume that the positions produced by the end of input function have a special origin, ``created out of nothing''.}.

When given an input string, the transducer works as follows. It begins in the initial state with all registers containing the empty string. Then it processes each input letter from left to right, updating the control state and the registers according to the transition function. Once the whole input has been processed, the end of input function is applied to the last state, yielding another sequence of register operations,  and finally the value of the transducer is extracted from the output register.  For the origin semantics,   the origin of an output letter is the input letter over which that output letter was first created, using an operation of the second type.

 Observe that the register operations do not allow copying registers, this is an important restriction which guarantees, among other things, that the size of the input is linear in the size of the output.
		
\begin{ourexample}
	 By composing the atomic register  operations and  using additional registers, we can recover additional register operations such as ``add letter $b$ to the end  of register $r$'', ``add letter $b$ to the beginning  of register $r$'', ``move register $r$ to register $s$, leaving $r$ empty''. In the examples, we will use the additional operations. 
	
	Consider the function $w \mapsto ww^R$, where $w^R$ is the reverse of $w$. The transducer has no control (i.e.~one control state) and two registers, which are used to store $w$, $w^R$. 
	When it reads an input letter $a$, the transducer adds $a$ as to the end of the register storing $w$ and adds $a$ to the beginning of the register storing $w^R$. The end of input update concatenates both registers, and puts the result in the first register, which is the output register.
	
	A transducer for the duplication function is obtained in a similar way. Observe that since the register operations do not allow copying,  it is still necessary to have two registers, both storing $w$.
	\end{ourexample}

\paragraph*{Deterministic two-way automaton with output.} A deterministic two-way automaton with output is like a deterministic two-way finite automaton, except that every transition is additionally labelled by a string (possibly empty) over the output alphabet. The output of the automaton is the concatenation  of the strings labelling the transitions in the run, in the order that they were executed. The origin of an output letter is the input letter in the source of the transition that output it. Therefore,  we require that every output letter is produced for transitions that have their source in input letters, and not the  markers $\vdash$ and $\dashv$ that guard the input from both sides.

One can consider a nondeterministic version, i.e.~a  nondeterministic two-way automaton with output. To make the output a function of the input, we require the nondeterministic automaton to have a unique run over every input.  This relaxation does not add to the expressive power, as long as two-way automata are considered. On the other hand, if we consider nondeterministic one-way automata with output, then the model looses expressive power, and will be studied in Theorem~\ref{thm:order-preserving-characterisation}.

\paragraph*{MSO transduction.}
A string over an alphabet $A$ can be treated as a relational structure, whose universe is the positions of the string, and which has a binary  position order predicate $x < y$ and label predicates $a(x)$ for the letters of the alphabet. By evaluating formulas on such a structure, one can use formulas to define a string language.
As shown by independently by B\"uchi, Trakhtenbrot and Elgot,  regular languages are exactly those definable in mso.

To transform strings into strings, we can use mso interpretations.
An \emph{mso interpretation}  is a function from structures over some fixed  input  vocabulary to structures over some fixed  output  vocabulary, which is specified by a system of  mso formulas, as follows.  There is a  {universe formula} with one free  variable over the input vocabulary, which selects  the elements of the universe of the  input structure that will appear in the universe of the output structure. Furthermore,  for every  predicate  of the output vocabulary there is a  formula over the input vocabulary of the same arity, which says how the predicates are defined in the output structure.

Another function from structures to structures is called \emph{$k$-copying}; which maps a structure to  $k$ disjoint copies of itself, together with  unary predicates $1(x),\ldots,k(x)$ which   identify which copy an element comes from. 
A \emph{copying mso transduction} consists of first a copying function, followed by an mso interpretation.

Using the encoding of strings as relational structures, copying mso transductions can be used to map strings to strings.
Such a string to string transducer is called \emph{mso definable}. The origin information in such a transducer is defined in the natural way.

\paragraph*{Equivalence of the models.} Deterministic two-way automata with output are shown to defined the same transducers as copying mso transductions in~\cite{DBLP:journals/tocl/EngelfrietH01}. The same proof works if the semantics with origin information are used.  The streaming transducers are shown to be equivalent to the previous two models  in~\cite{DBLP:conf/fsttcs/AlurC10}; the same proof also works with the origin semantics.  A transducer with origin  is called \emph{regular} if it can be defined by any one of the three models mentioned above.

%% file: characterisation.tex
\section{A machine independent characterisation}
In this section we present a Myhill-Nerode style characterisation of regular transducers with origin information. Suppose that $f$ is a regular transducer with origin information, and $u,v,w$ are input strings. Some part of the output $f(uvw)$ is a transformed version of the middle part of the input $v$. Because of  regularity, the way that $v$ is transformed will depend only on some regular properties of the left and right parts of the input $u$ and $w$. In particular, there should be finitely many types of the left and right parts, as far as the transformation on the middle part is concerned. To see the way the middle part is transformed in the output, without looking at the artefacts of the left and right parts, we use the notation $f(\underline u| v | \underline w)$, which represents the output $f(uvw)$, with the parts originating from the left part $u$  and the right part $w$ abstracted away. For instance,  if $f$ is the duplicating function, then 
\begin{align}\label{eq:example-duplicating-function}
	f(ab|cd|e) = \leftvar \ \middleblock{cd}\ \rightvar \ \leftvar\ \middleblock{cd}\ \rightvar.
\end{align}
This notion is defined in more detail below.

\subsection{Derivatives}

	\paragraph*{Factored outputs.}
Let $X$ be a set of \emph{source identifiers}, which are supposed to correspond to the parts in a factorsiation of the input, e.g.~the source identifiers  can be ``left'' and ``right'' when the input is divided into a left and right factor.  An \emph{$X$-factored output}  is defined to be a sequence of pairs of the form (element from $X$, string over the output alphabet). The second coordinate may be undefined. Each pair is called a \emph{block}, the first coordinate of a block is called its \emph{source identifier}, and the second coordinate is called its \emph{output assignment}.  We write a factored output by putting the source identifiers  in a lower row, and the values of the output assignment in an upper row. For instance,  in the equation~\eqref{eq:example-duplicating-function}, the right side is an $X$-factored output, with $X$ being the set $\set{\leftvar,\middlevar,\rightvar}$.

Suppose that $w$ is an input, $\sigma$ is a colouring of the  positions in the input string,  and  $f$ is a   transducer with origin information. Factorise the output $f(w)$    so that consecutive  blocks of  letters with the same colour of their source are put in the same factor; and label each such block by the colour of their source.  This results in a factored output, which we denote by $f(w)/\sigma$. Define
\begin{align*}
f(u|v|w) \eqdef f(uvw)/\sigma
\end{align*}
where $\sigma$ maps positions from $u,v,w$ to ``left'', ``middle'' and ``right'', respectively. Define $f(v|w)$ in the similar way, only without ``middle'', which means that $f(v|w)=f(v|\epsilon|w)$.
 For example, if $f$ is the duplicating function $w \mapsto ww$, then 
\begin{align*}
	f(abc|d) =  \leftblock{abc}\ \rightblock d\ \leftblock{abc}  \ \rightblock d 
\end{align*}
 	If we underline some of the strings,   then we make the output assignment undefined for the blocks coming from the underlined strings, e.g.
\begin{align*}
	f(abc|\underline d) =  \leftblock{abc}\ \rightvar \ \leftblock{abc}  \ \rightvar.
\end{align*}
	\begin{ourexample}
		Suppose that $f$ is the function $w \mapsto ww$. Then 
		\begin{eqnarray*}
			f(a|b|c) &=& \overset a \leftvar \ \overset b \middlevar \ \overset c \rightvar \ \overset a \leftvar \ \overset b \middlevar \ \overset c \rightvar  \\
							f(\underline a|b|\underline c) &=&  \leftvar \ \overset b \middlevar \  \rightvar \  \leftvar \ \overset b \middlevar \  \rightvar \\
										f(a||bc) &=& \overset a \leftvar \  \rightvar \ \overset a \leftvar \  \rightvar  
		\end{eqnarray*}
	\end{ourexample}

	\paragraph*{Derivative.}	A two-sided derivative of a function $f$ is any function of the form
	\begin{align*}
		\mderiv u w \qquad \eqdef \qquad  v \mapsto f(\underline u| v | \underline w).
	\end{align*}
	  \emph{Left derivatives} and \emph{right derivatives} are   functions of the, respective, forms
	\begin{align*}
		\lderiv v  \qquad \eqdef \qquad  w \mapsto f(\underline v |w) \\
				\rderiv w  \qquad \eqdef \qquad  v \mapsto f( v  | \underline w).
	\end{align*}

	\begin{ourexample}
		Let $f$ be the function $w \mapsto w^Rw$. Then 
		\begin{align*}
		\mderiv u w (v) = \rightvar \ \middleblock {v^R} \ \leftvar\  \middleblock v\ \rightvar
		\end{align*}
		for every nonempty strings $u$ or $w$. When the string $u$ is empty, then the left block disappears, likewise when $w$ is empty then the right blocks disappear. In particular, this function has four possible values for the two-sided derivative. There are two possible values for the left derivative $\lderiv u$, namely the functions
		\begin{align*}
v \mapsto \rightblock {v^R v} \qquad\qquad v \mapsto \rightblock {v^R}\ \leftvar \ \rightblock {v}.
		\end{align*}
		The first function behaves like the orginal function, although they are technically distinct.
	\end{ourexample}
	
	\begin{ourexample}
		Let $f$ be the function which is the identity on strings of even length, and which erases strings of odd length.  This function has three possible left derivatives $\lderiv v$, depending on whether $v$ is empty, nonempty and even legnth,  or odd length. In the last case, the derivative is		
		\begin{align*}
			w \mapsto \begin{cases}
				\leftvar\ \rightblock w &\mbox{if $w$ has odd length} \\
				\epsilon & \mbox{otherwise}
			\end{cases}.
		\end{align*}
	\end{ourexample}

	\begin{ourexample}
		Consider an input alphabet $\set{a,b}$ and the function $f$ defined by 
		\begin{align*}
			f(a^nbw) = w^n \qquad f(a^n)=\epsilon.
		\end{align*}
This function has infinitely many left derivatives. For instance,   $\lderiv {a^nb}$ is
\begin{align*}
v \mapsto \rightblock{v^n}.
\end{align*}
	\end{ourexample}

\begin{ourexample}\label{ex:right-fin-left-infin} Here is  a function with finitely many right derivatives, but infinitely many left derivatives.
Consider first the  function which scans its input from left to right, and outputs only those letters whose position is a prime number
	\begin{align*}
		f(a_1 \cdots a_n) = w_1 \cdots w_{n} \qquad \mbox{where $w_i=$}\begin{cases}
			a_i & \mbox{if $i$ is a prime number}\\
			\epsilon & \mbox{otherwise}.
		\end{cases}
	\end{align*}
	This particular function has infinitely many right derivatives, since 
	\begin{align*}
		 \rderiv w (v) = \begin{cases}
		  \leftblock{f(v)}\ \rightblock	& \mbox{if there is a prime number in $\set{|v|+1,\ldots,|vw|}$} \\
		  		  \leftblock{f(v)}	& \mbox{otherwise}.
		 \end{cases}
	\end{align*}
	However finitely many right derivatives can be obtained by making the last position to be output unconditionally, i.e.~in the transducer
	\begin{align*}
		g(a_1 \cdots a_n) = f(a_1\cdots a_{n-1})a_n.
	\end{align*}
	In this case, $g$ has only two right derivatives, namely
	\begin{align*}
		v \mapsto \leftblock {g(v)} \qquad \qquad v \mapsto \leftblock {f(v)}\ \rightvar.
	\end{align*}
	The function has infinitely many left derivatives $\lgderiv v$ because the criterion ``$i$ is a prime number'' needs to be replaced by ``$i+|v|$ is a prime number''.
\end{ourexample}

\paragraph*{The characterisation. } A function   from strings to a finite set 
is called a \emph{regular colouring} if every preimage $f^{-1}(x)$ is a regular language. A function from tuples of strings to a finite set is called  a regular colouring  if it factors through 
a function \begin{align*}
	(w_1,\ldots,w_n) \mapsto (f_1(w_1),\ldots,f_n(w_n)) 
\end{align*}
for some regular colourings $f_1,\ldots,f_n$ of individual strings. In an equivalent definition, all the colourings $f_i$ could be required to be the same. In yet another definition, for every $x$ in the image, the set
\begin{align*}
	\set{w_1\#w_2\#\cdots\#w_n : f(w_1,\ldots,w_n)=x}
\end{align*}
is a regular language of strings, assuming $\#$ is a symbol not in the input alphabet. 

In the following theorem, a string-to-string function with origin information is any function which maps a string (over a fixed input alphabet) to another string (over a fixed ouptut alphabet), together with origin information. The theorem characterises those functions which happen to be regular transducers, e.g.~can be implemented by an mso transduction.

%

\begin{theorem}[Machine indpendent characterisation]\label{thm:myhill}
	For a string-to-string function with origin information, the following conditions  are equivalent
	\begin{enumerate}
		\item $f$ is regular transducer with origin information;
		\item  $f$ has finitely many left derivatives  and right derivatives;
		\item for every $a$ in the input alphabet, the following is a  regular colouring
		\begin{align*}
			(v,w) \mapsto f(\underline v | a | \underline w).
		\end{align*}
	\end{enumerate}
\end{theorem} 
The function in the third item of the theorem is called the \emph{characteristic function} of the transducer.  As we will see in the proof of the theorem, a transducer with origin information can be uniquely reconstructed based on its characteristic function. Therefore, instead of studying transducers, we can  study their characteristic functions. This is the case in the learning algorithm from Section~\ref{sec:learning}, and the studies of subclasses of transducers in Sections~\ref{sec:subclasses} and~\ref{sec:fo}.

As shown in Example~\ref{ex:right-fin-left-infin}, it is not enough to  require finitely many  derivatives of one kind, say right derivatives, since a function might have finitely many derivatives of one kind, but infinitely many derivatives of the other kind\footnote{ It does  follow from the theorem  that a function with finitely many left and right derivatives  has finitely many two-sided derivatives. This is because every regular string-to-string function has finitely many two-sided derivatives.}.

\paragraph*{Implication from 1 to 2.}
To show that there are finitely many derivatives in a regular transducer, suppose that $f$ is a function that is recognised by a two-way deterministic automaton with output.  When the head of this automaton enters a suffix of the input from the left in some state, then several things can happen:  it might not return from that suffix, or it might return from  that suffix  in some other state. In either case, whether it returns or not, the automaton can output an empty or nonempty string.  Define the \emph{suffix type} of a string to be the partial function 
	\begin{align*}
		Q \to (Q \cup \set{noreturn}) \times \set{empty,nonempty},
	\end{align*}
	which says what happens for each state $q$.  It is not difficult to see that $f(x| \underline y)$ depends only on the right type of $x$. Since there are finitely many suffix types, there are also  finitely many right derivatives. For the left derivatives, a similar argument works, only with a type that also describes the first time that a prefix is exited by the head.

\paragraph*{Implication from 2 to 3.}

	\begin{lemma}\label{lem:derivative-substitute} Let $w_1,\ldots,w_n$ and $w,v$ be input strings. Then
		\begin{align*}
			\lderiv {w} = \lderiv {v} \qquad \mbox{implies}\qquad f(\underline{w}|w_1| \cdots | w_n)= f(\underline{v}|w_1| \cdots | w_n)
		\end{align*}		
	\end{lemma}
	\begin{pr}
		By using the origin information, the value 
		\begin{align*}
			f(\underline{w}|w_1| \cdots | w_n)
		\end{align*}
		can be obtained from the value
		\begin{align*}
			f (\underline{w}|w_1 \cdots  w_n) \eqdef \lderiv w (w_1 \cdots  w_n).
		\end{align*}
	\end{pr}
	
	\begin{lemma}\label{lem:derivative-congruence}
		The functions $w \mapsto \lderiv w$ and $w \mapsto \rderiv v$ are regular colourings.
	\end{lemma}
	\begin{pr}
		The domains of the two functions are finite by the assumption 2, which says that there are finitely many left and right derivatives.
		By symmetry, we only study the left derivatives $\lderiv w$. By Lemma~\ref{lem:derivative-substitute}, it follows that 
		\begin{align*}
			\lderiv w = \lderiv v \qquad \mbox{implies} \qquad f(\underline w | a | u)= f(\underline v | a | u)
		\end{align*}
		hods for every input letter $a$ and every input string $u$.
		Since $f(\underline w | a | u)$ uniquely determines $f(\underline {wa} | u)$, it follows that 
		\begin{align*}
			\lderiv w = \lderiv v \qquad \mbox{implies} \qquad \lderiv {w a} = \lderiv {w a}
		\end{align*}
		holds for every input letter $a$.
		This means that  the set of left derivatives can be equipped with a transition function as in a deterministic left-to-right automaton, so that after reading a string $w$ from the state $\lderiv \epsilon$, the automaton ends up in state $\lderiv w$. 
	\end{pr}

		Thanks to Lemma~\ref{lem:derivative-substitute} and its symmetric version for right derivatives, 
		\begin{align*}
			(v,w) \mapsto f(\underline v | a | \underline w)
		\end{align*}
		factors through the function
		\begin{align*}
			(v,w) \mapsto (\lderiv v,\rderiv w),
		\end{align*}
		meaning that equal results for the second function imply equal results for the first function. The second function  is a regular colouring by Lemma~\ref{lem:derivative-congruence}. A function which factors through a regular colouring must itself be a regular colouring, which finishes the proof of item 3 in Theorem~\ref{thm:myhill}.
		
\paragraph*{Implication from 3 to 1.}  	In the proof, we use a two-way model called a \emph{lookaround transducer}, which is defined as follows.  
The control is given by two deterministic automata: a \emph{past automaton}, which is left-to-right deterministic, and a \emph{future automaton}, which is right-to-left deterministic. There is a set of \emph{registers}, with a designated \emph{output register}. The registers are updated by a \emph{register update function}, which inputs a state of the past automaton, an input letter, and a state of the future automaton, and outputs a sequence of register operations. 
		
		The output of the  transducer on a string  over the input alphabet is defined as follows.
		Define the \emph{type} of a position $i$ in the input string to be:  the state of the past automaton after doing a left-to-right pass over the prefix that ends just before position $i$; the label of position $i$; the state of the future automaton after doing a right-to-left pass over the suffix that begins just after position $i$. To each type, the register update function assigns a register update.  The automaton begins with all register empty, and then it executes the register updates corresponding to the types of all positions in the string, read from left to right.  After all of these register updates are executed, the value of the function is found in the  output register.
		
		Lookaround transducers define exactly the regular transducers, also under the origin semantics.

To prove item 1 in Theorem~\ref{thm:myhill}, we  construct a  lookaround  transducer based on the assumption that the chara. 	We begin by describing the registers. The transducer has a register for every left block  in every partial output  $f(\underline x| \underline y)$. A partial output $f(v|\underline w)$ can be interpreted as a register valuation, which is defined on  the left blocks of $f(\underline v|\underline w)$  and   undefined on all other registers. The transducer is   designed to satisfy the following invariant:  when it has finished processing a prefix $v$ of an input $vw$, then its register valuation is $f(v|\underline w)$.

It remains to define the past and future automata, as well as the register update function. Our assumption, namely item 3, says that for every letter $a$ in the input alpahbet, the function
		\begin{align*}
			(v,w) \mapsto f(\underline v | a | \underline w)
		\end{align*}
is a regular colouring. This means that there is a left-to-right deterministic automaton, which we can choose to be the past automaton, and a right-to-left deterministic automaton, which we can choose to be the future automaton, such that $f(\underline v | a | \underline w)$ depends only on the state of the past automaton after reading $v$, the input letter $a$, and the state of the future automaton after reading $w$ from right-to-left.    The following lemma shows that the register udpate function can be defined to satisfy the invariant.
\begin{lemma}\label{lem:}
	Based on $f(\underline v | a | \underline w)$, one can construct a sequence of register operations which transforms the register valuation corresponding to $f( v |\underline{aw})$ into the register valuation corresponding to  $f(va | \underline w)$

\end{lemma}
		\begin{pr} For every  left block $b$ in $f(va| \underline w)$ there is  a corresponding sequence $b'$ of left and middle blocks in  $f(\underline v | \underline a | \underline w)$. The register update required by the lemma is defined according to this sequence: for every left block $b$ in $f({va}|\underline w)$, its value is defined to be the concatenation, according to the sequence $b'$, of the values of the left and middle blocks in $f(v|\underline{aw})$ and $f(\underline v| a | \underline w)$, respectively. 	\end{pr}
	
	This finishes the proof of Theorem~\ref{thm:myhill}.

%% file: learning.tex
\section{Learning}
\label{sec:learning}
This section shows that transducers with origin information can be learned. We first recall the Angluin algorithm for regular languages, which will be used as a black box in our learning algorithm for learning transducers.
The setup for  the Angluin algorithm is as follows. A teacher knows a regular language. A learner  wants to learn  this language, by  asking two kinds of queries. In a \emph{membership query},  the learner gives a string and the teacher responds whether this string is in the language. In an \emph{equivalence query}, the learner proposes candidate for the teacher's  language, and the teacher either says that this candidate is correct, in which case the interaction is over, or returns a \emph{counterexample}, which is a string in  the symmetric difference between the candidate and teacher's languages.

Angluin proposed an algorithm~\cite{DBLP:journals/iandc/Angluin87}, in which the learner learns the language by asking a number of queries which is polynomial in the minimal automaton for the teacher's language, and the size of the counterexamples given during the interaction. In this section, we propose a variant of this algorithm, but for learning transducers with origin information. In the case of transducers, the membership query  becomes a \emph{value query}, where the learner gives a string and the teacher responds with the value of the transducer on that string.  In the equivalence query, the counterexample becomes a string  where the transducer proposed by the learner gives a different value than the transducer of the teacher. Both in the value query and in the counterexample, the teacher also provides the origin information.

\begin{theorem}\label{thm:learning}
	A regular string-to-string transducer with origin information can be learned using value and equivalence queries in polynomial time (both number of queries and computation time) in terms of the number of left and right derivatives, and the size of the counterexamples given by the teacher.
\end{theorem}
\begin{pr}
	By Theorem~\ref{thm:myhill}, learning a transducer with origin information $f$ is the same as learning the characteristic function
	\begin{align*}
		(v,w) \mapsto f(\underline v | a | \underline w).
	\end{align*}
We  show that the characteristic function can be encoded as a regular language, and then the  Angluin algorithm can be invoked as a black box to learn it.  A value of the characteristic function can be seen as a string over the output alphabet, plus two additional letters $\leftvar$ and $\rightvar$ for indicating left and right blocks. Therefore, the characteristic function can be interpreted as a string language
	\begin{align*}
		L_f  \eqdef \set{v \# a \#w^R \# f(\underline v | a | \underline w)},
	\end{align*}
	where the strings before the first and second $\#$ are over the input alphabet, and the string after the second $\#$ is over the output alphabet extended by the letters for left and right blocks. It is not difficult to see that the minimal deterministic automaton of the language $L_f$ is polynomial in the parameters from the statement of the lemma. The reason why the string $w$ is reversed is that the automaton for the right derivatives reads its input from right to left.
	
	Therfore, we can apply the Angluin algorithm to learn the language $L_f$. The only technical issue is that the queries for learning $L_f$ need to be translated into the queries for learning $f$. A membership query
	\begin{align*}
		u \stackrel ? \in L_f
	\end{align*} corresponds to a value query for the transducer $f$, as follows. If the string $w$ does not have the right format -- exactly three apperances of $\#$, with exactly one letter between the second and third appearance  --  then the learner can immediately answer ``no'' without bothering the teacher. Otherwise, the learner extracts the $(v,a,w)$ stored in $u$, and asks for the value of $f(vaw)$. Using the origin information, learner computes the value $f(\underline v | a | \underline w)$, and can thus determine if $u$ belongs to $L_f$.  The correspondence between equivalence queries and counterexamples is done in a similar fashion.
\end{pr}


%% file: subclasses.tex
\section{Order-preserving transducers}
\label{sec:subclasses}
In this section, we present two characterisations of sublcasses of transducers.
For the standard semantics without origins,~\cite{DBLP:conf/lics/FiliotGRS13} shows how to decide if a deterministic two-way transducer is equivalent to a nondeterministic one-way transducer, while~\cite{DBLP:conf/stacs/WeberK94} shows how to decide (in polynomial time) if a nondeterministic one-way transducer is equivalent to a deterministic one-way transducer. This sections shows  analogous results for the origin semantics. Unlike~\cite{DBLP:conf/lics/FiliotGRS13,DBLP:conf/stacs/WeberK94}, the characterisations for the origin semantics are self-evident,  which shows how origin information makes some technical problems go away.  A more difficult characterisation, about first-order definable transducers, is presented in the next section.

\begin{theorem}\label{thm:order-preserving-characterisation}
	For a regular string-to-string transducer with origin information $f$, the following conditions are equivalent.
	\begin{enumerate}
		\item\label{item:order-preserving} for every input string $w$, the origin mapping from positions in the output $f(w)$ to the input $w$ is order-preserving. 
		\item\label{item:left-right-division}  $f(\underline v | \underline w)$ belongs to $(\epsilon +\leftvar)(\epsilon + \rightvar)$ for every input strings $v,w$.
		\item\label{item:one-register-lookahead}  $f$ is recognised by a  streaming transducer with lookahead which has only one register, and which only appends output letters to that register.
		\item\label{item:unique-nondeterministic}  $f$ is recognised by a  nondeterministic one-way automtaton with output, which has exactly one run over every input string.
	\end{enumerate}
\end{theorem}
\begin{pr}
	The implication from item~\ref{item:order-preserving} to item~\ref{item:left-right-division} follows straight from the definition. 
	For the implication from item~\ref{item:left-right-division}  to~\ref{item:one-register-lookahead}, we  observe that  if condition 1 is satisfied, then the transducer constructed in the proof of Theorem~\ref{thm:myhill} will only have one register, and it will only append letters to that register during the run. For the implication from item~\ref{item:one-register-lookahead}  to item~\ref{item:unique-nondeterministic}, we observe that a nondeterministic one-way automaton with output can guess, for each position of the input, what the lookahead will say. Since the lookahead is computed by a deterministic right-to-left automaton, this leads to a unique run on every input string. The implication from item~\ref{item:unique-nondeterministic} to item~\ref{item:order-preserving} also follows straight from the definition.
\end{pr}

Observe that the condition in item~\ref{item:left-right-division} can be effectively decided, even in polynomial time, when the characteristic function of the transducer is known.

We can  further restrict the model by requiring that the transducer in item~\ref{item:one-register-lookahead} does not use any lookahead, or equivalently, by requiring that the automaton in item~\ref{item:unique-nondeterministic} be deterministic. This restricted model is characterised in the following theorem.

\begin{theorem}\label{thm:deterministic-without-lookahead}
	Let $f$ be a regular transducer which satisfies any of the equivalent conditions in Theorem~\ref{thm:order-preserving-characterisation}. Then $f$ is defined by a one-way deterministic automaton with output if and only if every input strings $u,v,w$ satisfy
	\begin{align*}
		f(u|\underline v) = f(u|\underline w)
	\end{align*}
\end{theorem}
\begin{pr}
	The left-to-right implication is immediate. For the right-to-left implication, we observe that the assumption implies that
	\begin{align*}
		f(\underline u | a | \underline v)
	\end{align*}
	does not depend on $v$, but only on $\lderiv u$ and the letter $a$. Furthermore, since $f$ satisfies the assumptions from Theorem~\ref{thm:order-preserving-characterisation}, the above value is of the form
	\begin{align*}
		\leftvar \ \rightblock{x},
	\end{align*}
	where $x$ is a possibly empty string over the output alphabet, and the block $\leftvar$ is possibly missing. After reading input $u$, the automaton stores in its control state the derivative $\lderiv u$. When it reads a letter $a$, it updates its control state, and outputs the string $w$, which depends only on the control state and input letter $a$.
\end{pr}


%% file: fo.tex
\section{First-order definable transducers}
\label{sec:fo}
A regular string-to-string transducer with origin information is called \emph{first-order definable} if it can be defined by an mso transduction which does not use set quantification, but only the first-order quantifiers. In this section, we characterise the first-order definable transducers by a variant of aperiodicity.

A language is called \emph{first-order definable} if there is a sentence of first-order logic that is true in strings from the language and false for other strings. This lifts, in the natural way, to a notion of a first-order definable regular coloring. %

\begin{theorem}\label{thm:first-order transducer}
	The following conditions are equivalent for  a regular string-to-string transducer $f$ with origin information.
	\begin{enumerate}
		\item\label{item:fo-transduction} it is definable by a first-order string-to-string transduction.
		\item\label{item:fo-derivatives} the colorings $w \mapsto \lderiv w$ and $w \mapsto \rderiv w$ are first-order definable.
		\item\label{item:fo-five} for every letters $a,b$, the following  is a  first-order definable coloring
		\begin{align*}
			(u,v,w) \mapsto f(\underline u | a | \underline v | b| \underline w)
		\end{align*}
			\end{enumerate}
\end{theorem}
Before proving the theorem, we observe that condition in item~\ref{item:fo-derivatives} is effective. 
Using a straightforward extension of the  the Schutzenberger-McNaughton-Papert theorem, one can effectively decide if a regular coloring is first-order definable. By applying the decision procedure to the functions $w \mapsto \lderiv w$ and $w \mapsto \rderiv w$, we can decide if a regular transducer is first-order definable.

Without origin information, a variant of first-order definable transducers was considered in~\cite{DBLP:journals/jcss/McKenzieSTV06}, namely the transducers which are first-order definable in the sense of Theorem~\ref{thm:first-order transducer} and  simultaneously order preserving in the sense of Theorem~\ref{thm:order-preserving-characterisation}.    For instance, the doubling transduction $w \mapsto ww$ is first-order definable in the sense of Theorem~\ref{thm:first-order transducer}, but not in the sense of~\cite{DBLP:journals/jcss/McKenzieSTV06}, because it is not order preserving.
By testing for both condition~\ref{item:fo-derivatives} from Theorem~\ref{thm:first-order transducer} and condition~\ref{item:left-right-division} of Theorem~\ref{thm:order-preserving-characterisation}, we get an effective characterisation of the transducers from~\cite{DBLP:journals/jcss/McKenzieSTV06}, however  this characterisation uses origin information

\paragraph*{Implication from~\ref{item:fo-transduction} to~\ref{item:fo-derivatives}.} 
By symmetry, it suffices to show that  $w \mapsto \lderiv w$  is  first-order definable.
Since there are finitely many possible values of $\lderiv w$, there is a finite set of test strings such that $\lderiv w$ is uniquely determined by its values on the test strings. It is therefore sufficient to show that for every  test string, one can compute in first-order logic, given $w$, the value of $\lderiv w$ on the test string. This is done in the following lemma.
\begin{lemma}\label{lem:two-partition-fo-definable}
	If $f$  is first-order definable, then for every string $v$ the function $w \mapsto f(\underline w|v)$
	is first-order definable.
\end{lemma}
\begin{pr}
	Define the \emph{colored version} of an alphabet to be two disjoint copies: one called the black version, and one called the red version. Define $f'$ to be a transducer which  inputs a string over the colored version of the input alphabet, and works the same way as $f$, except that it outputs strings over the colored version of the output alphabet, and the color of an output letter is inherited from the corresponding input letter. Consider a function which inputs a string $w$ over the original input alphabet, and outputs a black version of $w$, followed by a red version of $v$. We denote this function by $w \mapsto w\markalpha{v}$.    It is not difficult to see that both functions described above are first-order definable transducers, and since these are closed under composition, it follows that $w \mapsto f'(w\markalpha{v})$ is first-order definable.   For every possible value $x$ of $f(\underline w|v)$, one can write a first-order formula $\varphi_x$ such that 
  \begin{align*} 
  	f( \underline w |v ) = x \quad \mbox{iff} \quad f'(w\markalpha{v}) \models \varphi_x \qquad \mbox{for every $w$}.
  \end{align*}
  The property on the right side of the equivalence can be checked by a first-order formula working on $w$. 
\end{pr}


\paragraph*{Implication from~\ref{item:fo-five} to~\ref{item:fo-transduction}.} For the moment, we skip the implication from~\ref{item:fo-derivatives} to~\ref{item:fo-five}, which is the most difficult\footnote{The implication from~\ref{item:fo-transduction} to~\ref{item:fo-five} can be proved in the same way as in  Lemma~\ref{lem:two-partition-fo-definable}. Therefore the theorem with only items~\ref{item:fo-transduction} and~\ref{item:fo-five} would be easier to prove. Such a weaker theorem would still give an effective characterisation of first-order definable transduction.}.
We begin by showing that the characteristic function of the transducer is first-order definable.
\begin{lemma}\label{lem:first-oder-definable-characteristic-function}
	For evey letter $a$, the following  function is first-order definable
	\begin{align*}
	(v,w) \mapsto f(\underline v | a | \underline w).
	\end{align*}
\end{lemma}
\begin{pr}
	The for every letter $b$ of the input alphabet,   function
	\begin{align*}
		(v,w') \mapsto f(\underline v | a | \underline {w'b})
	\end{align*}
	must be first-order definable since it 
	factors through the function
	\begin{align*}
		(v,w') \mapsto f(\underline v |a| \underline {w'} |b| \underline \epsilon ),
	\end{align*}
	which is first-order definable. It follows that, as long as we know that  $w$ ends with  $b$, then we can use first-order logic to obtain $f(\underline v | a | \underline w)$. If $w$ is nonempty, then it has finitely many possiblities for the last letter, and therefore, we can use first-order logic to obtainw $f(\underline v | a | \underline w)$ as long as we know that $w$ is nonempty. The same argument works when $v$ is nonempty. The remaining case is when both $v$ and $w$ are empty, which can be detected in first-order logic.
\end{pr}

We are now ready to define the first-order transduction. Define the \emph{contribution} of a position in an input to  be the subsequence (a string) of the output  which  originates from that position. The contribution depends only on   $f(\underline {v}  | a | \underline {w})$, where $v$ is the part of the input  before the position, $a$ is the label of the position, and $w$ is the part of $w$ after the position.  
 
Let  $N$ be the maximal length of a contribution, ranging over all finitely many choices of  $f(\underline v|a|\underline w)$. 
The first-order interpretation defining $f$ will copy each position of the input at most $N$ times.

Given an input $a_1 \cdots a_n$, the fist-order interpretation works as follows.
\begin{itemize}
	\item {\bf The universe formula.}  For $i \in \set{1,\ldots,N}$, the 
	$i$-th copy of a position $x$ in the input string belongs to the universe of the output string if and only if $i$ is at most the length of the contribution of the $x$-th letter in the input string. For fixed $i$, this can be determined by a first-order formula with a free variable $x$, thanks to Lemma~\ref{lem:first-oder-definable-characteristic-function}.
	
	%
%
		\item {\bf The label formulas.}  The label formulas are defined in the same way as the domain formula, only using the label of the $i$-th contribution.
 	\item {\bf The order formula.}  For every $i,j \in \set{1,\ldots,N}$, we need a first-order formula with two free variables $x$ and $y$ which says if, in the output, the $i$-th letter in the contribution of position $x$ comes before the $j$-th letter in the contribution of position $y$. Assuming that $x$ is before $y$, this information is entirely determined by  partial output \begin{align*}
	f(\underline{a_1 \cdots a_{x-1}}|a_x|\underline{a_{x+1}  \cdots a_{y-1}} | a_y | \underline{a_{y+1} \cdots a_n}),
\end{align*}
which can be obtained using first-order logic thanks to the assumption. The case when $x$ comes after $y$ is symmetric, and in the special case when $x=y$ the formula simply returns the value of $i \le j$.
\end{itemize}

\paragraph*{Implication from~\ref{item:fo-derivatives} to~\ref{item:fo-five}.} 
We say that a regular coloring   $g$ of $n$-tuples   is \emph{aperiodic on coordinate $i$} if for every  strings
	\begin{align*}
		w_1,w_2,\ldots,w_{i-1}, x,y,z, w_{i+1},\ldots,w_n,
	\end{align*}
	the following function is ultimately constant
	\begin{align*}
		i \mapsto  g(w_1,\ldots,w_{i-1},xy^iz, w_{i+1},\ldots,w_n).
	\end{align*}
A straightforward 	consequence of the Sch\"utzenberger-McNaughton-Papert theorem is that
	a regular colouring is first-order definable if and only if it is aperiodic on every coordinate.
Therefore,  item~\ref{item:fo-five} will follow once we show that the function
\begin{align*}
	(u,v,w) \mapsto (\underline u | a | \underline v | b|  \underline w)
\end{align*}
is aperiodic   on every coordinate.  We begin with the first coordinate. Let then  $u_1,u_2,u_3,v,w$ be strings.  We need to show that  the function
	\begin{align*}
		i \mapsto 	f(\underline {u_1u_2^iu_3} |a| \underline {v}|b | \underline {w})
	\end{align*}
	is ultimately constant.  For fixed $a,v,b,w$, the function above  factors through 
	\begin{align*}
		i \mapsto \lderiv {u_1 u_2^i u_3},
	\end{align*}
	which must be ultimately constant by the assumption in item 3, and therefore it must also be ultimately constant. A symmetric argument shows that the function from the statement of the lemma is  aperiodic on the third coordinate.
	We are  left with  the  second coordinate. Let then $u,v_1,v_2,v_3,w$ be strings. We need to show that the function
	\begin{align}
		\label{eq:aperio}
		i \mapsto 	f(\underline u |a| \underline{v_1 v_2^i v_3}|b | \underline w)
	\end{align}
is ultimately constant.

\begin{lemma}\label{lem:double} Suppose that 
	\begin{align*}
		f(\underline{uv}|\underline{w}) = f(\underline{u}|\underline{w}) = f(\underline{u}|\underline{vw})
	\end{align*}
	Then the function
	\begin{align*}
		i \mapsto f(\underline u | \underline{v^i} | \underline w)
	\end{align*}
	is ultimately constant.
\end{lemma}

Before proving the lemma, we show how it completes the proof of the implication from item~\ref{item:fo-derivatives} to item~\ref{item:fo-five} in  Theorem~\ref{thm:first-order transducer}, and therefore also completes the proof of the theorem itself. Our goal is to show that the function~\eqref{eq:aperio} is ultimately constant. We will show that for sufficiently large $j$, the function
	\begin{align}
		\label{eq:aperio-bis}
		i \mapsto 	f(\underline u |a| \underline{v_1v_2^j}|\underline{v_2^i}|\underline{v_2^jv_3}|b | \underline w)
	\end{align}
is ultimately constant. This will imply that the function
\begin{align*}
	i \mapsto f(\underline u |a| \underline{v_1 v_2^{i+2j} v_3}|b | \underline w)
\end{align*}
is ultimately constant, and therefore also~\eqref{eq:aperio} is ultimately constant.
We claim that the function~\eqref{eq:aperio-bis} factors through the following functions
\begin{eqnarray}
i & \mapsto &	f(\underline {u}|a|\underline{v_1v_2^j}| \underline{v_2^{i+j}v_3bw})\label{eq:one-two-of-five}\\
i & \mapsto &	f(\underline {uav_1v_2^j} | \underline{v_2^{i}}| \underline {v_2^jv_3bw})\label{eq:three-of-five}\\
i & \mapsto &	f(\underline {uav_1v_2^{j+i}}| v_2^jv_3|b|w)\label{eq:four-five-of-five}.
\end{eqnarray}
Indeed, the value of~\eqref{eq:aperio-bis} is obtained from~\eqref{eq:three-of-five} by replacing the left part with the first three parts in~\eqref{eq:one-two-of-five}, and replacing the right part with the last three parts in~\eqref{eq:four-five-of-five}. The function~\eqref{eq:one-two-of-five} factors through 
\begin{align*}
	i \mapsto \rderiv{v_2^{i+j}v_3bw}
\end{align*}
and is therefore ultimately constant by the assumption that $w \mapsto \rderiv w$ is aperiodic. For the same reason, the function~\eqref{eq:four-five-of-five} is ultimately constant. We are only left with showing that~\eqref{eq:three-of-five} is ultimately constant.   If $j$ is large enough, then by aperiodicity of $w \mapsto \lderiv w$, we see that
\begin{align*}
	\lderiv {uav_1v_2^j} = 	\lderiv {uav_1v_2^{j+1}} \qquad \mbox{and} \qquad \rderiv{v_2^jv_3bw}=\rderiv{v_2^{j+1}v_3bw}
\end{align*}
which implies that the assumptions of Lemma~\ref{lem:double} for 
\begin{align*}
	u = 
	uav_1v_2^j \qquad v = v_2 \qquad w = v_2^jv_3bw.
\end{align*}
The conclusion of the lemma shows that~\eqref{eq:three-of-five} is ultimately constant.

\begin{pr}[of Lemma~\ref{lem:double}]
	Consider the partial output  $f(\underline u|\underline w)$, which consists of left and right blocks in alternation. This output can be viewed as a graph (which consists of a single directed path), call it $G_0$, which is illustrated in the following picture. 
	\begin{center}
		\includegraphics[scale=0.3]{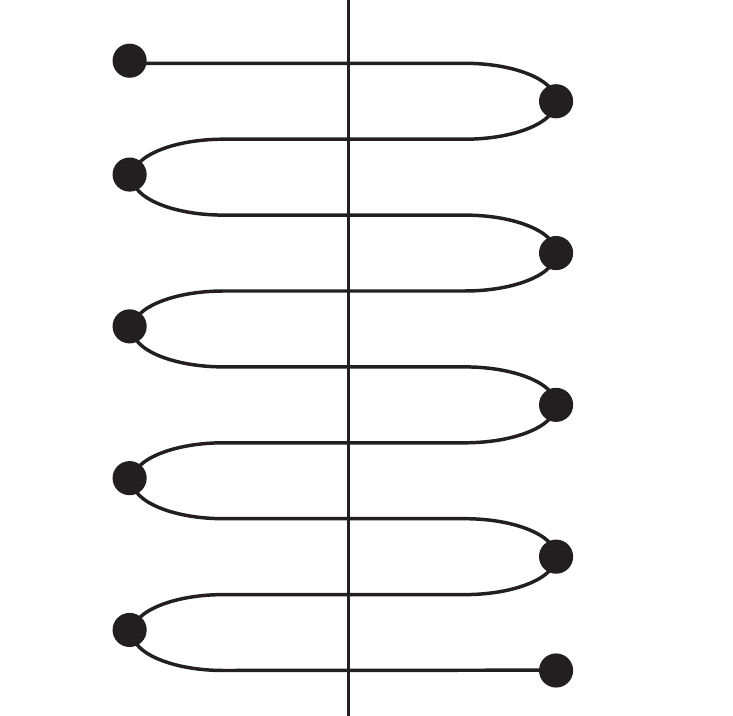}
	\end{center}
	The vertices (black dots in the picture)  in the left column correspond to left blocks, the vertices in the right column correspond to right blocks. 	There is a directed edge from a block to the following block; the edges in the picture are implicitly directed so that the path goes from top to bottom.
		Now consider the graph, call it $G_1$, which corresponds to the partial output $f(\underline u| \underline v | \underline w)$, which is illustrated in the following picture.
	\begin{center}
		\includegraphics[scale=0.3]{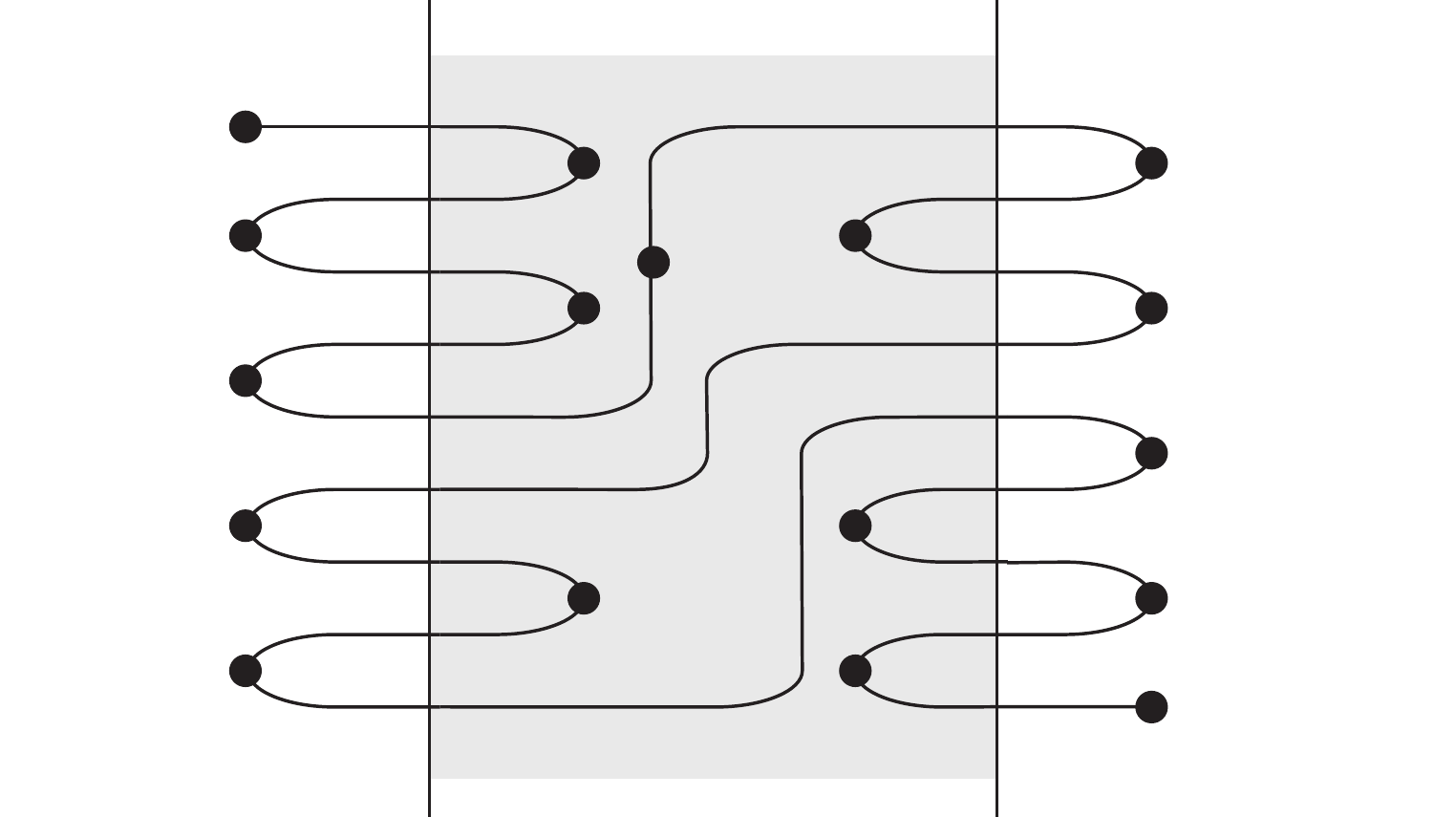}
	\end{center}
	The middle blocks are the vertices in  the grey area. Thanks to the assumption
	\begin{align*}
		f(\underline{u}|\underline{vw}) = f(\underline{u}|\underline{w}),
	\end{align*}
 the left blocks are visited in the same order as in $G_0$. Using the other assumption, the right blocks are visited in the same order as in $G_0$.  However, the combined order on both left and right blocks might be different in $G_0$ and $G_1$.
	
 Finally, consider the graph, call it $G_i$,  which corresponds to
 \begin{align*}
 	f(\underline u| \overbrace{ \underline v  | \underline v | \cdots | \underline v |}^{\mbox{$i$ times}}  \underline w),
 \end{align*} which is illustrated below for $i=5$. 
	\begin{center}
		\includegraphics[scale=0.3]{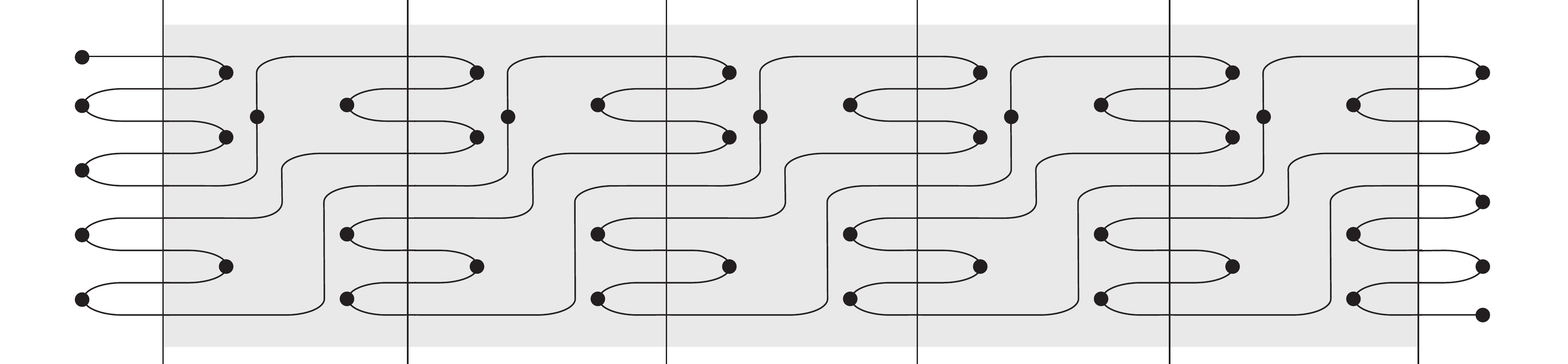}
	\end{center}
	Notice that the nodes in the left part of the graph, which correspond to the left blocks, are the same in every graph $G_i$, the same holds for the right blocks; also the order on left blocks and the order on right blocks are the same. 
	The middle blocks  of $f(\underline u| \underline v^i | \underline w)$  correspond to maximal paths which are entirely contained in the grey area. For a left or right block $x$ and a number $i$, consider a vertex $\sigma_i(x)$ and a boolean value $\tau_i(x)$, defined as follows.
	\begin{itemize} 
				\item $\sigma_i(x)$  is the first left or right block after  $x$ in the graph $G_i$ (if it exists).
				\item $\tau_i(x)$ says if  the path in $G_i$ from $x$ to $\sigma_i(x)$ passes through a vertex in the grey area.
	\end{itemize}
	The sequence of blocks in $f(\underline u| \underline v^i | \underline w)$  consists of the left and right blocks listed according to the function $\sigma_i$, with a middle block appended after those blocks $x$ for which $\tau_i(x)$ says yes.  To prove the lemma, it therefore suffices to show that for large enough $i$,  the function $\sigma_i$ is always the same, likewise for $\tau_i$.

Define  $loop_i$ to be the same as $\sigma_i$, but with its domain restricted to blocks $x$ such that $x$ and $\sigma_i(x)$ have the same type (meaning both are left blocks, or both are right blocks).
\begin{lemma}\label{lem:}
	The function $loop_i$ is the same for large enough $i$, likewise for  $\tau_i$.
\end{lemma}
\begin{pr}
If  in the graph $G_i$ there is a path which connects two  blocks on the same type and does not pass through blocks of the other type, then same path is present in $G_j$. This means that
 if $i < j$ and $loop_i(x)$ is  defined, then $loop_i(x)=loop_j(x)$. 	Therefore, the functions $loop_i$ stabisilse eventually. A similar argument works for $\tau_i$.
\end{pr}
The following lemma finishes the proof.
\begin{lemma}\label{lem:}
	The function $\sigma_i$ can be uniquely determined from  $loop_i$.
\end{lemma}
\begin{pr}	
			As we have observed, the ordering on left blocks does not depend on $i$, likewise for right blocks.  In othcolourords, there is a unique ordering $<_L$ on left blocks and a unique ordering $<_R$ on right blocks.
The function $\sigma_i$ is a successor function (i.e.~a function that maps every element, except one, to a successor so that a linear order is formed) which satisfies the following conditions:
	\begin{itemize}
		\item the ordering induced by $\sigma_i$ on left blocks is $<_L$.
				\item the ordering induced by $\sigma_i$ on right blocks is $<_R$.
		\item $\sigma_i$  extends $loop_i$.
		\item $\sigma - loop_i$ maps left blocks to right blocks and vice versa.
		\end{itemize}

		It is not difficult to see that for every $loop_i$, there is at most one such function.
\end{pr}
This compeletes the proof of Lemma~\ref{lem:double}.
\end{pr}

%% file: conclusion.tex
\section{Further work}
It seems possible that the  ideas  in this paper extend to mso-definable tree-to-tree transductions. Another direction for further study is the computational complexity of equivalence.  It is possible  that for some of the machine models, by  going from standard to  origin  semantics, one  can lower the complexity of equivalence testing.